\documentclass[proof]{WileyASNA-v1}

\articletype{Original Article}%

\received{}
\revised{}
\accepted{}

\hypersetup{hidelinks}

\begin{document}

\title{Cometary dust activity as a proxy for refractory organic material release
}

\author[]{Serhii Borysenko}

\authormark{BORYSENKO \textsc{}}

\address[]{\orgname{Main Astronomical Observatory of NAS of Ukraine}, \orgaddress{\state{Kyiv}, \country{Ukraine}}}

%\address[2]{\orgdiv{Org Division}, \orgname{Org Name}, %\orgaddress{\state{State name}, \country{Ukraine}}}

%\address[3]{\orgdiv{Org Division}, \orgname{Org Name}, \orgaddress{\state{State name}, \country{Australia}}}

\corres{Serhii Borysenko, Main Astronomical Observatory of NASU, Akademika Zabolotnoho Str. 27, 03143, Kyiv, Ukraine. \email{borisenk@mao.kiev.ua}}

\abstract{Cometary dust is an important reservoir of refractory organic matter and may contribute to the redistribution of carbon-rich material throughout planetary systems. We investigate whether the widely used dust-activity parameter \(Af\rho\) can serve as a first-order observational proxy for the release of refractory organic material. A sample of 27 active small bodies representing active asteroids, main-belt comets, quasi-Hilda comets, quasi-Trojan objects, Jupiter-family comets, Centaurs, and long-period comets was compiled from published photometric studies. Representative near-perihelion, phase-corrected \(A(0)f\rho\) values span more than three orders of magnitude, from a few centimetres for weakly active asteroids and main-belt comets to \(\sim10^4\) cm for the most active long-period comets. The large dispersion within individual dynamical classes shows that orbital classification alone does not determine the level of dust activity. Direct compositional measurements of cometary dust, including the high refractory-organic fraction measured for 67P/Churyumov–Gerasimenko, support the interpretation of dust activity as an indirect indicator of potential organic-material release. However, a universal conversion of \(Af\rho\) into dust mass-loss rate is not possible because of uncertainties in grain-size distribution, density, porosity, velocity, and optical properties. We therefore suggest that \(Af\rho\) may be used as a comparative, first-order proxy for refractory-organic material release.}


%\jnlcitation{\cname{%
%\author{Borysenko S.} 
%\author{B. Hoskins}, 
%\author{R. Lee}, 
%\author{G. Masato}, and 
%\author{T. Woollings}} 
%(\cyear{2026}), 
%\ctitle{Cometary dust activity as a proxy for refractory organic material release}, \cjournal{Astronomische Nachrichten}, \cvol{2026;00:1--10}.}}

%%\fundingInfo{Funding info text.}

\maketitle

%\footnotetext{\textbf{Abbreviations:} ANA, anti-nuclear antibodies; APC, %antigen-presenting cells; IRF, interferon regulatory factor}

\section{Introduction}\label{sec1}

Comets are traditionally considered among the most primitive bodies of the Solar System. Their nuclei preserve volatile and refractory materials that formed during the earliest stages of planetary system evolution and, in some cases, may contain matter inherited from the protosolar molecular cloud. Because of their relatively low degree of thermal processing, comets provide a unique opportunity to investigate the inventory of organic compounds, water, and mineral components that existed during the formation of the Solar System.
For several decades, cometary astrobiology has focused primarily on the role of comets as carriers of water and organic compounds to the early Earth ~\citep{Chyba1990, Thomas2006}. The discovery of numerous organic molecules in cometary comae, together with in situ measurements obtained during the Rosetta mission and laboratory analyses of returned samples from comet 81P/Wild 2, strengthened the hypothesis that cometary bodies may have contributed significantly to the inventory of prebiotic material available on terrestrial planets ~\citep{Altwegg2019, Brownlee2006, Sandford2006, Cordiner2021, Hanni2024}. 

The hypothesis that comets may have supplied a significant fraction of the volatile and organic inventory of the early Earth has been discussed for more than three decades. One of the first quantitative estimates of cometary delivery of organic compounds was presented by Chyba et al. (1990), who demonstrated that cometary impacts could provide substantial amounts of prebiotic material during the heavy bombardment epoch. Since then, spacecraft missions such as Stardust and Rosetta have transformed this hypothesis from a largely theoretical concept into one supported by direct compositional measurements of cometary nuclei and dust.

However, most investigations have concentrated on volatile species observed spectroscopically in cometary atmospheres, whereas considerably less attention has been devoted to the role of cometary dust as a carrier of complex organic matter. Refractory organic matter is physically associated with cometary dust particles rather than volatile organic molecules released into the gas phase. The term "refractory organic matter" refers to complex carbon-rich solid material physically associated with cometary dust particles, including macromolecular organic compounds, insoluble organic matter, and CHON-rich particles identified in cometary dust and interplanetary dust particles. This definition excludes volatile organic molecules released directly into the gas phase during sublimation. Since refractory organics are predominantly incorporated into dust grains, cometary dust activity provides a more appropriate observational proxy for their transport than measurements of gaseous species alone.   

Recent studies indicate that refractory dust particles may represent one of the most important reservoirs of carbon-rich material in cometary nuclei. Laboratory investigations of interplanetary dust particles, micrometeorites, and Stardust samples demonstrate that cometary dust contains abundant organic matter, including compounds that formed under low-temperature conditions in the outer protoplanetary disk or even before the formation of the Solar System ~\citep{Mumma2011, BockeleeMorvan2004}. 

Consequently, the investigation of dust activity provides an
alternative and potentially more direct approach to
understanding the transport of organic matter throughout
planetary systems ~\citep{Engrand2023, Raponi2020}.
Particularly important in this context are different cometary populations (from long-period comets to main-belt comets and active asteroids).
For example, quasi-Hilda comets and main-belt comets, ~\citep{Hsieh2006, Jewitt2015} unlike classical bright comets observed near perihelion, 
often exhibit weak or absent gaseous emissions
while maintaining measurable dust activity.
Such activity may occur at heliocentric distances where
water sublimation is inefficient, indicating the operation of
alternative physical mechanisms ~\citep{Meech2004, Guilbert2012}.
Because dust dominates the observable manifestations of
activity, these objects provide valuable laboratories for
investigating the production, transport, and preservation of
refractory organic material.

\section{Analysis and Discussion}\label{sec2}

%Nulla malesuada porttitor diam. Donec felis erat, congue non, volutpat at, %tincidunt tristique, libero. Vivamus viverra
%fermentum felis. Proin fermentum massa ac
%quam. Sed diam turpis, molestie vitae, placerat a, molestie nec, %leo~\cite{Bland2001} Maecenas lacinia. Nam ipsum ligula, eleifend
%at, accumsan nec, suscipit a, ipsum. Morbi blandit ligula feugiat magna. %Nunc eleifend consequat lorem. 

%Example for bibliography citations cite~\citep{Kosugi2009}, %cites~\cite{Stix2004}

\subsection{Cometary Dust as a Reservoir of Complex Organic Matter}

The recognition that cometary dust contains significant quantities of organic material represents one of the most important developments in modern planetary science. Early cometary models primarily emphasized volatile ices such as water, carbon monoxide, carbon dioxide, methane, and ammonia. Although these compounds remain essential constituents of cometary nuclei, recent investigations reveal that the refractory fraction of cometary matter contains a surprisingly rich inventory of carbon-bearing compounds.

Evidence for the organic richness of cometary dust comes from multiple independent sources. Samples returned by the Stardust mission from comet 81P/Wild 2 contain complex mixtures of silicates, sulfides, and carbonaceous material ~\citep{Brownlee2006, Sandford2006}. Similarly, chondritic porous interplanetary dust particles and ultracarbonaceous Antarctic micrometeorites exhibit characteristics consistent with a cometary origin and contain abundant refractory organic matter. The organic components often display isotopic anomalies, particularly enhanced deuterium abundances, suggesting formation in extremely cold environments.
The Rosetta mission further demonstrated that the nucleus of comet 67P/Churyumov–Gerasimenko is covered by a widespread organic-rich surface layer. Spectroscopic observations revealed the presence of complex macromolecular carbonaceous material distributed across large portions of the nucleus. Such findings suggest that organic matter is not confined to isolated regions but instead represents a fundamental component of cometary nuclei ~\citep{Altwegg2019}.

Direct compositional measurements provide an important constraint on the organic fraction of cometary dust. Using the COSIMA mass spectrometer on board Rosetta, Bardyn et al. (2017) analysed dust particles collected from comet 67P/Churyumov–Gerasimenko and found that they contain a very large fraction of carbon-rich refractory material. The particles were inferred to consist of approximately 45\% organic matter by mass and 55\% mineral material, with substantial particle-to-particle variability. This result demonstrates that refractory organics can constitute a major fraction of cometary dust and therefore supports the use of dust activity as a first-order indicator of the potential release of refractory organic material.

From an astrobiological perspective, the distinction between volatile molecules and refractory organic matter is critical. Volatile species are rapidly released and dispersed into space during sublimation, whereas refractory dust particles may survive for extended periods. These particles can travel large distances, become incorporated into meteoroid streams, and potentially deliver organic compounds to planetary surfaces. Consequently, dust may represent a more efficient long-term transport mechanism for prebiotic material than gaseous molecules.
The preservation of organic matter within dust grains also provides protection against ultraviolet radiation and other destructive processes. Complex organic molecules embedded within refractory particles may therefore survive conditions that would destroy isolated molecules in the gas phase. This protective effect increases the likelihood that chemically significant compounds can be transported through interplanetary space and eventually deposited onto planetary surfaces.
An additional consideration concerns the relationship between dust production and organic-matter transport. Because the majority of refractory organics are physically associated with dust particles, the intensity of dust activity may serve as a first-order indicator of the quantity of organic material released by a comet. This interpretation expands the significance of traditional photometric parameters and suggests new astrobiological applications for broadband cometary observations ~\citep{Duprat2010, Engrand2023, Sandford2006, Raponi2020}.

\subsection{Dust Activity Beyond the Water-Ice Sublimation Zone}

The occurrence of significant cometary activity at large heliocentric distances presents one of the most intriguing problems in cometary physics. Classical models assume that cometary activity is primarily driven by sublimation of water ice, which becomes efficient only at relatively small heliocentric distances. Nevertheless, numerous comets exhibit substantial activity beyond the traditional water-sublimation zone.

Observations of long-period comets demonstrated substantial dust activity still on quite significant heliocentric distances (more than 5 AU).
 Despite its large distance from the Sun, the comets displayed a well-developed
coma and measurable dust production. Photometric analysis
showed that such objects maintained significant activity levels
long before reaching the inner Solar System. Similar behavior
suggests that
dust release can occur under conditions where water sublimation alone is insufficient to explain the observed activity.
Several physical mechanisms may contribute to distant
activity. Sublimation of highly volatile species such as carbon
monoxide and carbon dioxide can drive dust ejection at large
heliocentric distances. Crystallization of amorphous water ice
may release trapped gases and provide additional energy for
particle lifting. Thermal fracturing, rotational instability, and
impact processes may also contribute to dust production in
specific cases. Main-belt comets provide particularly important examples of distant dust activity. These objects occupy
dynamically asteroidal orbits while exhibiting transient comet-
like behavior. Their activity often occurs near perihelion but
at heliocentric distances considerably larger than those typical of active Jupiter-family comets. The existence of such
objects demonstrates that volatile-rich material survives within
the asteroid belt and may become exposed through various
activation mechanisms. Quasi-Hilda comets represent another
valuable population. These objects occupy unstable orbits
associated with the Hilda region and frequently display weak
dust activity with little detectable gas emission. Their observational properties suggest mostly an intermediate character
between classical comets and active asteroids. Because dust
dominates their appearance, they provide useful targets for
investigating the relationship between dust production and the
preservation of primitive material ~\citep{Borysenko2019, Borysenko2020, Borysenko2022}.

The astrobiological significance of distant activity lies in the relatively gentle physical environment under which dust release occurs. Lower temperatures reduce thermal alteration of organic compounds, increasing the probability that primitive material remains chemically intact. Consequently, distant active comets may provide a more faithful representation of the original organic inventory incorporated during Solar System formation.

\subsection{Main-Belt Comets and Volatile-Rich Bodies in the Inner Solar System}

The discovery of main-belt comets has fundamentally changed our understanding of the distribution of volatile material in the Solar System. Prior to the identification of these objects, active comets were generally associated with the Kuiper Belt, the scattered disk, or the Oort Cloud. The observation of cometary activity among bodies residing within the main asteroid belt demonstrated that significant amounts of volatile material can survive much closer to the Sun than previously assumed ~\citep{Jewitt2015}.
Main-belt comets occupy dynamically asteroidal orbits characterized by Tisserand parameters with respect to Jupiter greater than three. Unlike classical comets, these objects are believed to have formed in situ within the asteroid belt or in nearby regions of the protoplanetary disk. Consequently, their volatile inventory may provide direct information about the distribution of water and organic material in the inner Solar System during planetary formation ~\citep{Hsieh2006}.

Several observational programs have revealed that many main-belt comets exhibit weak but persistent dust activity. Photometric investigations indicate that these objects often possess relatively small nuclei, low dust-production rates, and narrow dust tails. Nevertheless, even modest levels of activity imply the existence of subsurface volatile reservoirs capable of sustaining particle release over extended periods. The recent observations of  P/2023 JN16 (Lemmon), for example, demonstrate that a small object with a nuclear radius below approximately 0.5 km can remain active over a significant fraction of its orbit. Such activity is likely associated with the release of large, slowly moving particles from the surface ~\citep{Borysenko2025}. 

The astrobiological importance of main-belt comets extends beyond their present activity. If these bodies contain primitive volatile-rich material preserved since the early Solar System, they may represent relics of the population that contributed water and organics to the terrestrial planets. Unlike long-period comets originating in the distant Oort Cloud, main-belt comets occupy orbits that intersect the inner Solar System more frequently and at lower relative velocities. As a result, impacts involving such bodies may be more favorable for the survival of fragile organic compounds.
Another important aspect concerns the relationship between main-belt comets and active asteroids. Although some active asteroids display activity caused by rotational instability or impact events rather than sublimation, both classes contribute dust and potentially organic-rich material to interplanetary space. The boundaries between these populations remain uncertain, suggesting that a continuum may exist between volatile-rich and volatile-poor active bodies.
The existence of volatile-bearing objects within the asteroid belt also has implications for the formation of habitable planets in extrasolar systems. If volatile-rich minor bodies can survive in relatively warm regions around stars, then mechanisms analogous to those operating in the Solar System may be common elsewhere. Main-belt comets therefore provide valuable analogues for understanding the distribution of water and prebiotic material in planetary systems throughout the Galaxy ~\citep{Kwok2019, Anslow2023}.

\subsection{$A\!f\!\rho$ as a Proxy for Organic-Matter Transport}

The $A\!f\!\rho$ parameter introduced by A'Hearn and collaborators ~\citep{AHearn1984} has become one of the most widely used quantitative indicators of cometary dust activity. Traditionally, $A\!f\!\rho$ is interpreted as a measure of dust production and provides a convenient method for comparing activity levels among different comets and observing epochs. However, the parameter may also possess important astrobiological implications.

Physically, $A\!f\!\rho$ reflects the amount of sunlight scattered by dust particles within a specified photometric aperture. Larger $A\!f\!\rho$ values generally indicate greater quantities of dust in the coma and therefore stronger dust activity. Numerous observational studies of different cometary populations have demonstrated the usefulness of $A\!f\!\rho$ for characterizing the strength and evolution of cometary activity.
If cometary dust contains significant amounts of organic matter, as demonstrated by laboratory and spacecraft investigations, then $A\!f\!\rho$ may also serve as an indirect measure of organic-material release. Under this interpretation, a comet exhibiting strong dust activity is simultaneously injecting larger quantities of refractory carbonaceous matter into interplanetary space.

This concept provides a new framework for interpreting photometric observations. Instead of viewing $A\!f\!\rho$ exclusively as a dynamical or physical parameter, it becomes possible to regard it as a proxy for the efficiency of organic-material transport. Although the exact relationship between $A\!f\!\rho$ and organic mass flux depends on particle composition and size distribution, the general connection remains physically plausible.

Broadband photometric investigations of active bodies between 2 and 5 AU reveal substantial diversity in $A\!f\!\rho$ values. Some objects display only modest dust production, while others exhibit activity levels comparable to those of classical comets. Such differences may correspond to significant variations in the amount of organic material released into the Solar System environment ~\citep{Borysenko2020}.

Future studies combining photometric observations, spectroscopic measurements, and dust modeling could establish quantitative relationships between $A\!f\!\rho$ and organic-matter production rates. Such calibrations would transform a traditional observational parameter into a valuable astrobiological diagnostic, enabling systematic comparisons of the organic-transport efficiency among different cometary populations.

The observational studies of different cometary populations reveal substantial variations in dust productivity expressed through the $A\!f\!\rho$ parameter. Traditionally, $A\!f\!\rho$ has been interpreted as a measure of dust abundance in the coma and is widely used for comparing activity levels among different objects. However, recent spacecraft missions and laboratory investigations demonstrate that cometary dust contains a significant fraction of carbonaceous and organic material. 

We cannot directly derive the amount of refractory organic material from $A\!f\!\rho$, because: 1) $A\!f\!\rho$ measures dust scattering cross-section; 2) the fraction of organics in the dust is not directly observed; 3) the fraction of biologically relevant molecules is even less certain.
Consequently, $A\!f\!\rho$ may also serve as an indirect indicator of the amount of potentially prebiotic matter released into interplanetary space.
Assuming that a fraction $(f_{\rm org})$ of the total dust mass consists of organic compounds, the mass of organic material released by the comet may be expressed as
$$
M_{\rm org}=f_{\rm org}M_{\rm dust},
$$
where $M_{\rm org}$ -- mass of refractory organic material,
$M_{\rm dust}$ -- total dust mass.
In terms of production rates,
$$
\dot{M}_{\rm org}=f_{\rm org}\dot{M}_{\rm dust},
$$
where $\dot{M}_{\rm dust}$ -- dust mass production rate.
And the total amount of organic material released over a time interval \(t_1\)–\(t_2\) is

$$
M_{\rm org} =f_{\rm org}\int_{t_1}^{t_2}\dot{M}_{\rm dust}(t)dt
$$

The parameter \(Af\rho\), however, does not measure dust mass directly. It characterizes the effective scattering cross-section of dust within the photometric aperture and depends on the grain-size distribution, particle density, outflow velocity, porosity, albedo, and scattering properties. Therefore, a unique conversion between \(Af\rho\) and \(\dot{M}_{\rm dust}\), and consequently between \(Af\rho\) and \(\dot{M}_{\rm org}\), cannot be established without adopting a specific dust model. In the present study, \(Af\rho\) is therefore used only as a comparative, first-order indicator of dust activity and of the potential release of refractory organic material (Figure 1).

For comet 67P/Churyumov–Gerasimenko, COSIMA measurements suggest an organic mass fraction of approximately 45\% in cometary dust (50–1000 $\mu$m dust particles), although this value should not be assumed to be universal for all comet populations (Bardyn et al., 2017).

\begin{figure*}[t]
    \centering
    \includegraphics[width=\textwidth]{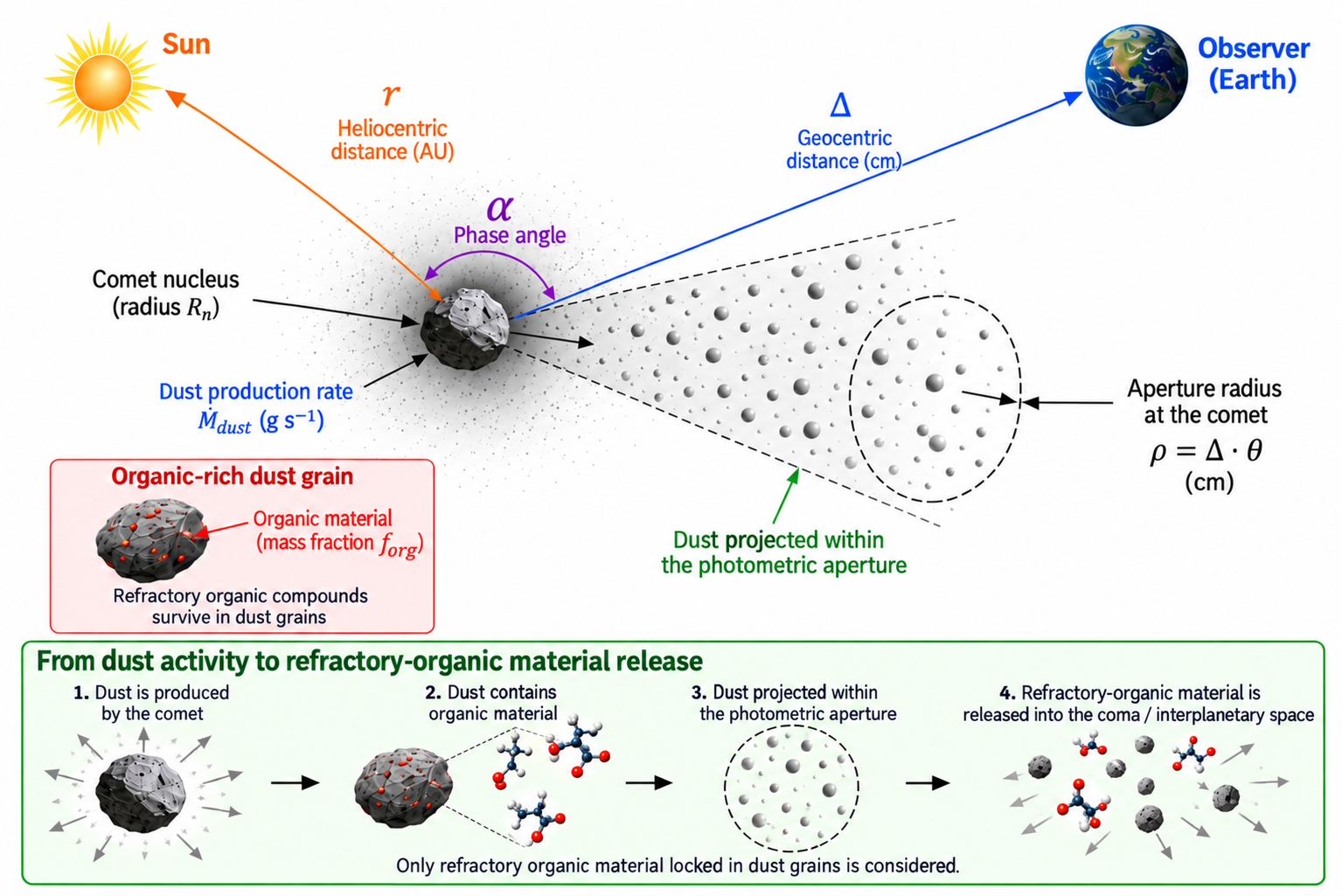}
    \caption{Conceptual relation between cometary dust activity and the release of refractory organic material.}
    \label{fig:1}
\end{figure*}

This relation suggests that $A\!f\!\rho$ can be interpreted not only as a measure of dust activity but also as a first-order proxy for the release of organic material. Under otherwise comparable dust properties, larger \(Af\rho\) values may indicate a larger potential release of refractory organic material.
The observations obtained for distant comets and active minor bodies indicate that significant dust production may occur even at heliocentric distances exceeding 3 -- 4 AU, where thermal processing remains relatively weak. Under such conditions, organic compounds embedded within dust particles are more likely to preserve their primordial chemical characteristics. As a result, distant active comets may represent efficient carriers of prebiotic matter and play an important role in the long-term redistribution of organic material between planetary environments (Table 1). The orbital elements \(q\), \(e\), and \(a\) correspond to the current osculating orbits, whereas the \(Af\rho(0)\) values refer to the observational epochs cited in the last column.

The sample was compiled from published photometric studies providing dust-activity measurements for representative members of the principal active-small-body populations considered here. Where required, the observed \(Af\rho(\alpha)\) values were corrected to zero phase angle according to \(Af\rho(0)=Af\rho(\alpha)/\Phi(\alpha)\). Near-perihelion measurements were preferentially selected to reduce differences associated with orbital phase. The sample is intended for comparative purposes and is not a statistically complete survey of the individual dynamical populations.

To compare the characteristic dust activity of different cometary populations, we used near-perihelion \(A(0)f\rho\) values, which minimize differences caused by sampling objects at widely different orbital phases. Figure 2 shows that the dust activity spans several orders of magnitude, from only a few centimetres for weakly active asteroids and main-belt comets to \(10^3\)–\(10^4\) cm for Jupiter-family, Centaur, and long-period comets. The large dispersion within individual dynamical classes indicates that orbital classification alone does not determine the level of dust production (Figure 2).

For the objects taken from Betzler et al. (2023), the published phase-corrected near-perihelion \(Af\rho(0)\) values were adopted directly without further modification. For objects analysed in our previous studies (Borysenko et al. 2019, 2020, 2022), the published \(Af\rho\) measurements were recalculated to zero phase angle according to

$$ Af\rho(0)=\frac{Af\rho(\alpha)}{\Phi(\alpha)}, $$

where \(\Phi(\alpha)\) is the adopted cometary dust phase function. Representative measurements obtained close to perihelion were selected. Thus, all dust-activity values listed in Table 1 are expressed on a common zero-phase-angle basis.

A direct conversion of \(A(0)f\rho\) into dust mass-loss rate was not attempted here. Such a conversion depends strongly on the assumed grain-size distribution, particle density, outflow velocity, porosity, and scattering properties, which differ among individual objects and dynamical populations. Therefore, in the present study \(A(0)f\rho\) is used only as a comparative measure of dust activity and, consequently, as a first-order indicator of the potential release of refractory organic material.

\begin{center}
\begin{table*}[t]%
\centering
\caption{Orbital parameters and representative near-perihelion dust activity of selected active small bodies.\label{tab1}}%
\tabcolsep=0pt%
\begin{tabular*}{40pc}{@{\extracolsep\fill}lcccll@{\extracolsep\fill}}
\toprule
\textbf{Objects}  & \textbf{q (au)}  & \textbf{e} & \textbf{a (au)} &\textbf{$A\!f\!\rho(0)$ (cm)}  &  \textbf{References} \\
\midrule
\textbf{Active Asteroids}   &   \\
311P/PANSTARRS & 1.94  & 0.12  & 2.19 & 2.6 $\pm$ 0.1 & Borysenko et al. 2020   \\
331P/Gibbs     & 2.88  & 0.04  & 3.00 & 4.8 $\pm$ 0.4 & Borysenko et al. 2020   \\  
\textbf{Main Belt Comets} & \\
133P/Elst -- Pizarro (2014) & 2.67 & 0.16 & 3.16 & 15 $\pm$ 2 & Borysenko et al. 2020 \\
238P/Read                   & 2.37 & 0.25 & 3.17 & 4.5 $\pm$ 0.4 & Borysenko et al. 2020 \\
288P/(300163)               & 2.44 & 0.20 & 3.05 & 9  $\pm$ 1 & Borysenko et al. 2020 \\
\textbf{Quasi Hilda Comets} &  \\
65P/Gunn                    & 2.93  & 0.25 & 3.89 & 187 $\pm$ 28 & Betzler et al. 2023 \\
117P/Helin -- Roman -- Alu 1 & 3.09 & 0.25 & 4.12 & 332 $\pm$ 24 & Betzler et al. 2023 \\
246P/NEAT                    & 3.40 & 0.21 & 4.32 & 2480 $\pm$ 41 & Betzler et al. 2023 \\
362P/(457175)                & 2.87 & 0.28 & 3.97 & 88   $\pm$ 3 & Borysenko et al. 2019 \\
\textbf{Quasi Trojan Comets}  & \\
74P/Smirnova -- Chernykh      & 4.84 & 0.05 & 5.11 & 139 $\pm$ 16 & Betzler et al. 2023 \\
244P/Scotti                   & 3.93 & 0.20 & 4.91 & 96 $\pm$ 11 & Betzler et al. 2023 \\
P/2019 LD2 (ATLAS)            & 4.47 & 0.13 & 5.16 & 280 $\pm$ 10 & Borysenko et al. 2022 \\
\textbf{Jupiter Family Comets} & \\
17P/Holmes                     & 2.09 & 0.43 & 3.64 & 605 $\pm$ 27 & Betzler et al. 2023 \\
19P/Borrelly                    & 1.31 & 0.64 & 3.61 & 1641 $\pm$ 241 & Betzler et al. 2023 \\
21P/Giacobini -- Zinner        & 1.01 & 0.71 & 3.49 & 1849 $\pm$ 41 & Betzler et al. 2023 \\
37P/Forbes                     & 1.62 & 0.53 & 3.46 & 513 $\pm$ 9   & Betzler et al. 2023 \\
78P/Gehrels                    & 2.00 & 0.46 & 3.73 & 768 & Betzler et al. 2023 \\
237P/LINEAR                    & 2.00 & 0.43 &   3.52 & 2045 $\pm$ 181 & Betzler et al. 2023 \\
\textbf{Trans Jupiter Comets (Centaurs)} & \\
29P/Schwassmann -- Wachmann 1    & 5.80 & 0.04 & 6.06 & 2631 $\pm$ 62 & Betzler et al. 2023 \\
\textbf{Long Periodic Comets} & \\
C/2009 P1 (Garradd)           & 1.53 & 0.999923 & -- & 10837 $\pm$ 26 & Betzler et al. 2023 \\
C/2011 L4 (PANSTARRS)         & 0.30 & 0.999914 & -- & 13214 $\pm$ 12 & Betzler et al. 2023 \\
C/2012 F6 (Lemmon)            & 0.74 & 0.998303 & -- & 9394 $\pm$ 709 & Betzler et al. 2023 \\
C/2012 X1 (LINEAR)            & 1.62 & 0.989857 & -- & 8486 $\pm$ 54 & Betzler et al. 2023 \\        
C/2013 R1 (Lovejoy)           & 0.80 & 0.998112 & -- & 8977 $\pm$ 35 & Betzler et al. 2023 \\
C/2013 X1 (PANSTARRS)         & 1.31 & 1.000984 & -- & 5235 $\pm$ 48 & Betzler et al. 2023 \\
C/2014 Q2 (Lovejoy)           & 1.29 &  0.996671 & -- &  19243 $\pm$ 113 & Betzler et al. 2023 \\
C/2014 S2 (PANSTARRS)         & 2.06 &  0.988218 & -- & 9636 $\pm$ 70 & Betzler et al. 2023 \\
\bottomrule
\end{tabular*}
\end{table*}
\end{center}

\begin{figure*}[t]
    \centering
    \includegraphics[width=\textwidth]{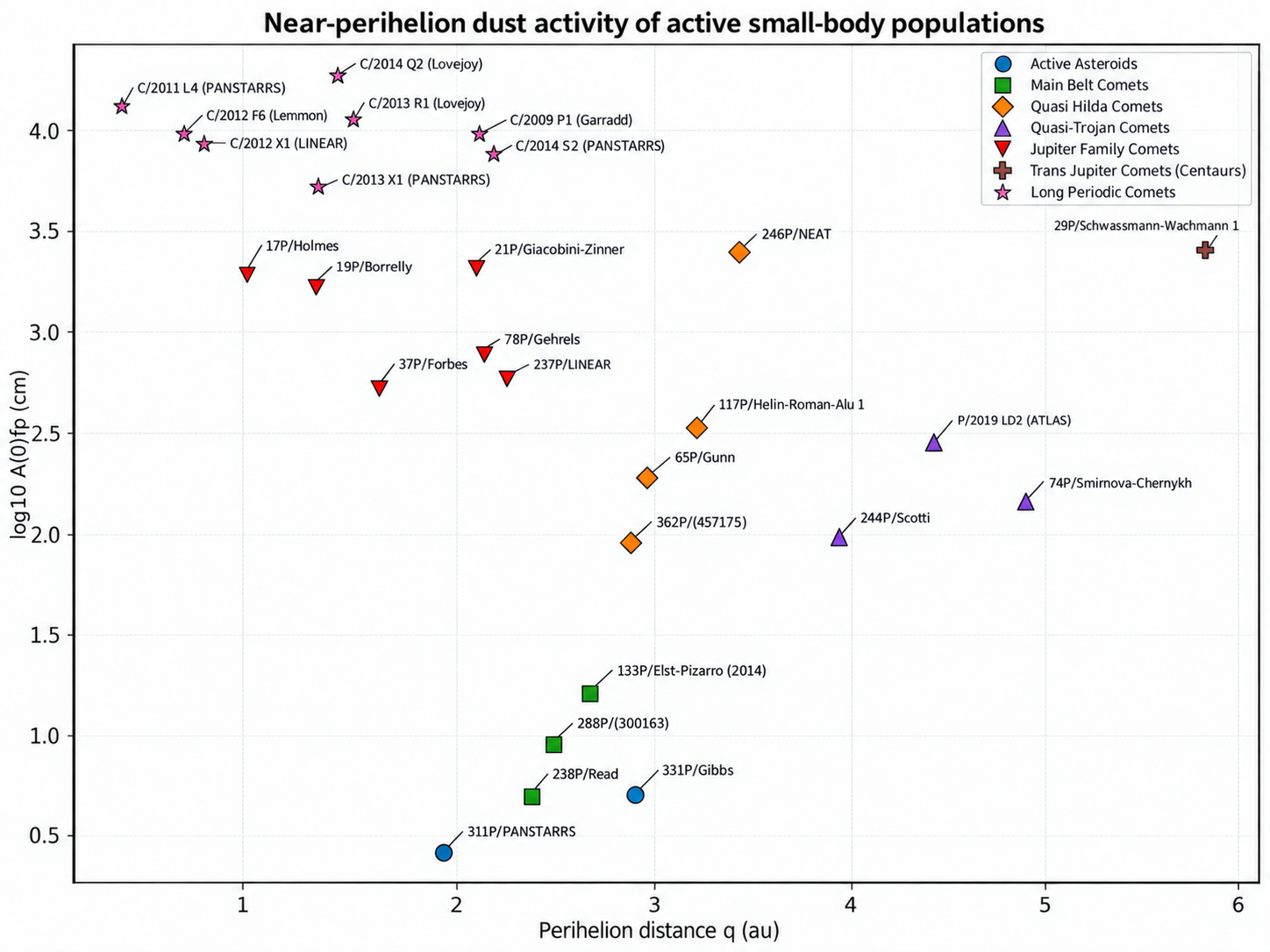}
    \caption{Near-perihelion dust activity of selected active small bodies as a function of perihelion distance. Dust activity is represented by the phase-corrected parameter \(A(0)f\rho\); different dynamical populations are indicated by different symbols and colors.}
    \label{fig:2}
\end{figure*}

\begin{figure*}[t]
    \centering
    \includegraphics[width=\textwidth]{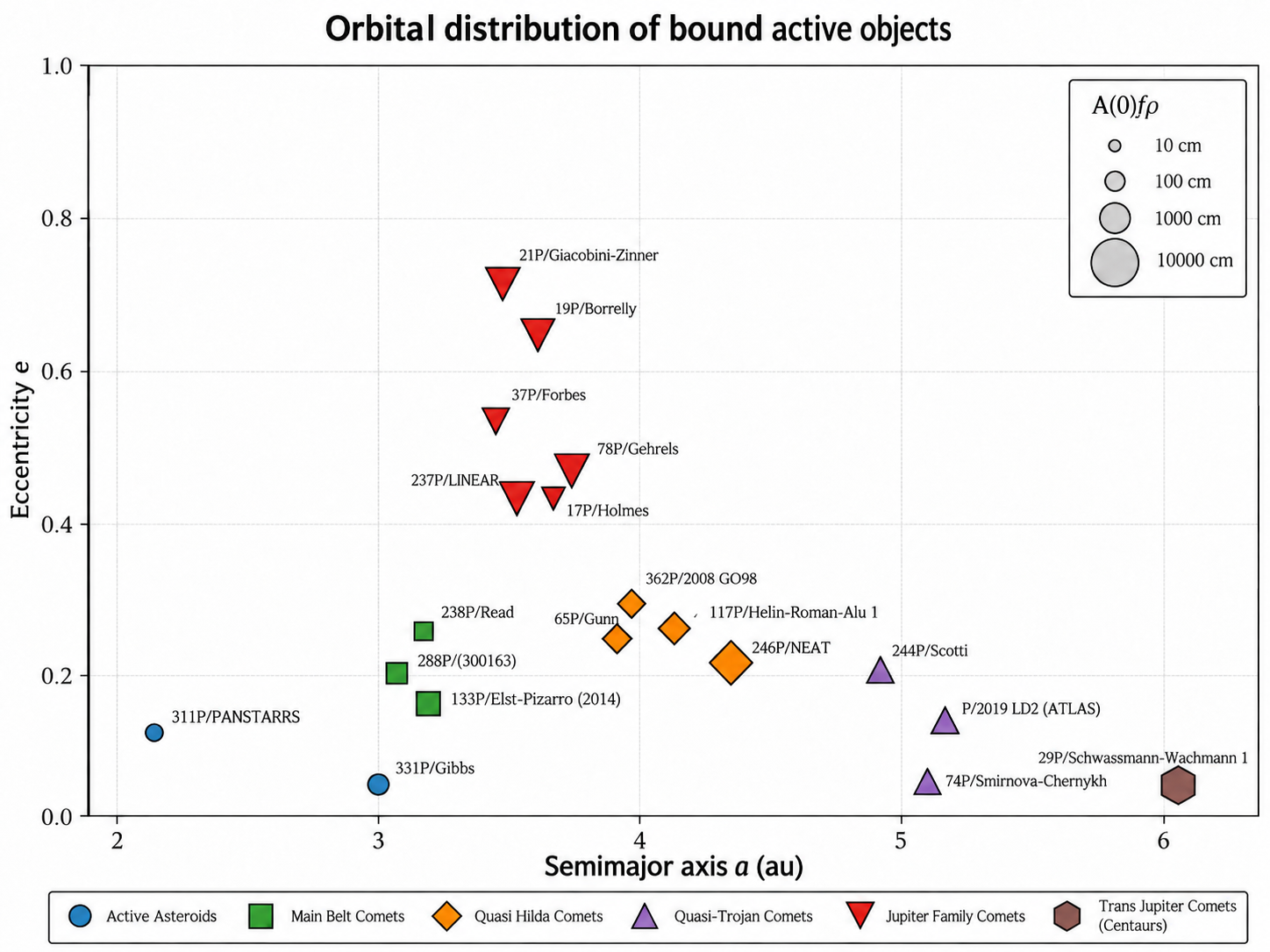}
    \caption{Distribution of selected active small bodies in the semimajor axis–eccentricity (\(a\)–\(e\)) plane. Different dynamical populations are indicated by color, while the symbol size is scaled according to the phase-corrected dust-activity parameter \(A(0)f\rho\). Long-period comets are not shown because their nearly parabolic orbits make the semimajor axis unsuitable for direct comparison on this scale.}
    \label{fig:3}
\end{figure*}

\subsection{Astrobiological Consequences for Early Earth and Exoplanets}

The presence of organic-rich dust in comets has important implications for theories concerning the origin of life. While the exact pathway leading from simple molecules to biological systems remains uncertain, it is widely accepted that prebiotic chemistry required substantial quantities of carbon-bearing compounds, water, and nitrogen-containing species. Comets are capable of supplying all of these ingredients.
Traditional discussions of cometary delivery have focused on large impact events occurring during the early history of the Earth ~\citep{Thomas2006, Chyba1990}. Such impacts undoubtedly contributed volatile and organic material. However, dust transport may have provided an additional and potentially more continuous delivery mechanism. Unlike catastrophic impacts, the accretion of cometary dust occurs gradually and can operate over billions of years.

The delivery of organic-rich particles to planetary atmospheres offers several potential advantages. Small particles experience less severe shock heating during atmospheric entry than larger bodies. Consequently, a greater fraction of their organic inventory may survive deposition onto planetary surfaces. Interplanetary dust particles collected in Earth's stratosphere demonstrate that extraterrestrial organic matter can indeed survive atmospheric passage.

Main-belt comets demonstrate that volatile-rich material can survive within the asteroid belt. Dynamical transport from this reservoir could therefore have contributed to the delivery of primitive material to the terrestrial-planet region, although the efficiency of such transport is not addressed here.

The astrobiological implications extend beyond the Solar System. Observations of protoplanetary disks and extrasolar planetary systems indicate that planet formation is accompanied by the production of large numbers of minor bodies. If these bodies contain organic-rich dust analogous to that observed in Solar System comets, then mechanisms of organic transport may operate universally.

Recent theoretical studies suggest that cometary impacts can deliver prebiotic molecules to rocky exoplanets under a wide range of conditions. The efficiency of delivery depends on planetary architecture, impact velocities, and stellar properties. Dust transport provides an additional pathway that may be less sensitive to these factors. Consequently, cometary dust may contribute to chemical enrichment even in systems where large impacts are relatively infrequent ~\citep{Anslow2023}.

From this perspective, cometary dust can be viewed as a universal carrier of molecular complexity. The widespread presence of carbon-rich particles in planetary systems implies that prebiotic ingredients may be distributed far more efficiently than previously assumed. Although such delivery does not guarantee the emergence of life, it increases the probability that suitable chemical environments can develop on habitable worlds ~\citep{Kwok2019, Engrand2023}.

\subsection{Limitations of the Proposed Proxy}
The approach proposed in this work should be regarded as a qualitative framework rather than a method for determining the absolute mass of organic material transported by comets. Although the dust-production parameter $A\!f\!\rho$ is widely used as a proxy for cometary dust activity, its conversion into the mass of refractory organic matter is currently limited by several poorly constrained parameters.

The principal uncertainty arises from the unknown fraction of refractory organic material contained within cometary dust particles. Measurements obtained by the Rosetta mission, laboratory analyses of Stardust samples, and studies of interplanetary dust particles demonstrate that cometary dust contains abundant carbon-rich organic matter. However, the relative abundance of refractory organics varies among different comets and remains insufficiently constrained for most comet populations. Consequently, a universal conversion factor between dust production and transported organic mass cannot presently be established.

Additional uncertainties are associated with the particle-size distribution, dust density, grain porosity, and the optical properties that influence the measured $A\!f\!\rho$ values. These parameters may differ substantially among long-period comets, Jupiter-family comets, quasi-Hilda comets, and main-belt comets. As a result, absolute estimates of transported organic mass would require assumptions that are presently difficult to justify observationally.

For these reasons, the present study does not attempt to derive absolute values of the transported refractory organic mass. Instead, $A\!f\!\rho$ is interpreted as a relative photometric indicator of the potential efficiency of refractory organic-material transport. Under the assumption that refractory organics are predominantly associated with dust particles, objects exhibiting stronger dust activity are expected to possess a greater capability for transporting organic matter through the Solar System. The proposed approach is therefore intended primarily for comparative studies of different comet populations rather than for determining absolute transport rates.

Future progress will depend on improved constraints on the composition of cometary dust obtained from sample-return missions, in situ spacecraft investigations, and laboratory analyses of cometary analogues. Such measurements may eventually permit calibration of the relationship between dust activity and refractory organic-matter transport, allowing the present qualitative framework to evolve into a quantitative diagnostic tool.

\section{Conclusions}\label{sec5}

Modern investigations of comets increasingly demonstrate that dust activity plays a central role in the transport and preservation of primitive Solar System material. While traditional cometary studies emphasized volatile species released into the coma, recent observations and laboratory analyses reveal that refractory dust constitutes a major reservoir of organic matter.
The evidence reviewed in this work suggests that cometary dust should be considered an important potential agent of prebiotic material. Organic-rich particles observed in cometary environments preserve chemical information inherited from the early Solar System and, potentially, from pre-solar interstellar environments. Their transport through interplanetary space provides a mechanism for distributing molecular complexity across planetary systems.
Particular attention has been devoted to main-belt comets, quasi-Hilda comets, and distant active comets. These populations exhibit dust-dominated activity under conditions where thermal processing is relatively weak. As a result, they may preserve primitive organic matter more effectively than comets observed close to the Sun. Their study therefore offers a unique opportunity to investigate the original inventory of carbonaceous material available during planet formation.

The $A\!f\!\rho$ parameter, traditionally used as a measure of dust production, may also possess astrobiological significance as an indirect indicator of organic-material release. Future studies combining photometry, spectroscopy, laboratory analyses, and dust-dynamical modeling could transform $A\!f\!\rho$ into a quantitative diagnostic of organic transport efficiency.

Finally, cometary dust transport provides a natural connection between planetary science and astrobiology. The continuous distribution of organic-rich particles throughout planetary systems may have contributed significantly to the chemical enrichment of the early Earth and may perform a similar role in extrasolar systems. In this sense, cometary dust is not merely a by-product of cometary activity but a potentially fundamental agent in the cosmic distribution of the ingredients required for life.

\nocite{*}
\bibliography{Wiley-ASNA}
\end{document}